\documentclass[onecolumn,amsmath,amssymb,aps,prl,superscriptaddress]{revtex4}
\usepackage{graphics}
\usepackage{newtxtext}
\usepackage{newtxmath}
\usepackage{natbib}
\usepackage{epstopdf}
\usepackage{amssymb}
\usepackage{xcolor}
\usepackage{float} 
\usepackage[caption = false]{subfig}
\usepackage{appendix}
\usepackage{graphicx}
\usepackage{amsmath}
\usepackage{color}
\usepackage{array}
\usepackage{subeqnarray}
\usepackage{ulem}
\usepackage[english]{babel}
\usepackage{setspace}
\usepackage{cleveref}
\usepackage{bm}
\begin{document}

\title{Soft-lubrication interactions between a rigid sphere and an elastic wall}
\author{Vincent Bertin}
\affiliation{Univ. Bordeaux, CNRS, LOMA, UMR 5798, 33405 Talence, France.}
\affiliation{UMR CNRS Gulliver 7083, ESPCI Paris, PSL Research University, 75005 Paris, France.}
\author{Yacine Amarouchene}
\affiliation{Univ. Bordeaux, CNRS, LOMA, UMR 5798, 33405 Talence, France.}
\author{Elie Rapha\"{e}l}
\affiliation{UMR CNRS Gulliver 7083, ESPCI Paris, PSL Research University, 75005 Paris, France.}
\author{Thomas Salez}
\email{thomas.salez@u-bordeaux.fr}
\affiliation{Univ. Bordeaux, CNRS, LOMA, UMR 5798, 33405 Talence, France.}
\affiliation{Global Station for Soft Matter, Global Institution for Collaborative Research and Education, Hokkaido University, Sapporo, Hokkaido 060-0808, Japan.}
\date{\today}

\begin{abstract}
The motion of an object within a viscous fluid and in the vicinity of a soft surface induces a hydrodynamic stress field that deforms the latter, thus modifying the boundary conditions of the flow. This results in elastohydrodynamic (EHD) interactions experienced by the particle. Here, we derive a soft-lubrication model, in order to compute  all the forces and torque applied on a rigid sphere that is free to translate and rotate near an elastic wall. We focus on the limit of small deformations of the surface with respect to the fluid-gap thickness, and perform a perturbation analysis in dimensionless compliance. The response is computed in the framework of linear elasticity, for planar elastic substrates in the limiting cases of thick and thin layers. The EHD forces are also obtained analytically using the Lorentz reciprocal theorem.
\end{abstract}

\date{}
\maketitle
\section{Introduction}
\label{sec:intro}

The fluid-structure interaction between flows and boundaries is a central situation in continuum mechanics, encountered at many length and velocity scales. A classical example is lubrication, where the addition of a liquid film, a lubricant, between two contacting objects, allows for a drastic reduction of the friction between them. Such a process occurs in a large variety of contexts with hard materials, such as roller bearings, pistons and gears in industry~\citep{dowson2014elasto}, or faults~\citep{brodsky2001elastohydrodynamic} and landslides~\citep{campbell1989self} in geological settings. At large velocity, or moderate loading, the liquid film is continuous with no direct contact between the solids. When the solids are deformable, the friction force can be described using elastohydrodynamic (EHD) models within the soft-lubrication approximation~\citep{dowson2014elasto}. 

The previous EHD coupling is also widely encountered in soft condensed matter, but at very different pressure and velocity scales~\citep{karan2018small}. Examples encompass the remarkable frictional properties of eyelids~\citep{jones2008elastohydrodynamics} and cartilaginous joints~\citep{mow1984fluid,jahn2016lubrication}, as well as biomimetic gels~\citep{gong2006friction} and rubbers~\citep{sekimoto1993mechanism,moyle2020enhancement,wu2020lubricated,hui2021friction}. Of interest as well are the collisions and rebounds of spheres in viscous environments~\citep{davis1986elastohydrodynamic,gondret2002bouncing,tan2019criterion}, the rheological properties of soft suspensions and pastes~\citep{sekimoto1993mechanism,meeker2004slip}, and the self-similar properties of the contact~\citep{Snoeijer2013}.  

In the last decade, EHD interactions have been of great interest in the material-science community with the emergence of contactless rheological methods to measure the mechanical properties of confined liquids and soft surfaces~\citep{chan2009dynamic,vakarelski2010dynamic,leroy2011hydrodynamic,leroy2012hydrodynamic,villey2013effect,wang2015out,wang2017elastic_b,wang2017elastic_a,guan2017noncontact,wang2018viscocapillary,laine2019micromegascope,bertin2020non}. Typically, in such experimental systems, a spherical colloidal probe is immersed in a fluid and driven to oscillate, with a nanometric amplitude, near a surface of interest. The force exerted on the probe is measured by an atomic force microscope, a surface force apparatus or a tuning-fork microscope, and depends on the properties of both the fluid and the solid boundary.

Generally, an object that moves in a confined fluid environment experiences an enhanced drag force with respect to the bulk Stokes law, as a result of the boundary-induced flow modification~\citep{happel2012low}. Furthermore, near a soft wall, the hydrodynamic interactions are modified by the deformation of the boundary that they generate, yielding a nonlinear coupling. Perturbation methods, assuming a small deformation of the interface, have been employed in order to calculate the soft-lubrication interactions exerted on a free infinite cylinder immersed in a viscous fluid and near a thin compressible elastic material~\citep{salez2015elastohydrodynamics}. In particular, interesting inertial-like features have been predicted despite the low-Re-number aspect of the flow.

Perhaps the most emblematic example of soft-lubrication interaction is the non-inertial lift force predicted for a particle sliding near a soft boundary~\citep{sekimoto1993mechanism,beaucourt2004optimal,skotheim2004soft,skotheim2005soft,urzay2007elastohydrodynamic,urzay2010asymptotic}. It might have important implications for advected biological entities, such as red blood cells~\citep{grandchamp2013lift} and vesicles~\citep{abkarian2002tank}. Only recently, the associated dynamical repulsion from an immersed soft interface has been studied experimentally. A preliminary qualitative observation was reported in the context of smart lubricants and elastic polyelectrolytes~\citep{Bouchet2015}. Then, a study involving the sliding of an immersed macroscopic cylinder along an inclined plane pre-coated with a thin layer of gel, showed quantitatively an effective reduction of friction induced by the EHD lift force~\citep{Saintyves2016}. Subsequently, the same effect was observed in the trajectories of micrometric spherical beads within a microfluidic channel coated with a biomimetic polymer layer~\citep{davies2018elastohydrodynamic}, and through the sedimentation of a macroscopic sphere near a pre-tensed suspended elastic membrane~\citep{rallabandi2018membrane}. Finally, direct measurements of the EHD lift force for two types of elastic materials have been performed at small scales, using surface force apparatus and atomic force microscopy, respectively~\citep{Vialar2019, zhang2020direct}. 

Despite the increasing number of EHD studies involving spherical probes, the soft-lubrication interactions of a free spherical object immersed in a viscous fluid and moving near an elastic substrate still have to be calculated. In the present article, we aim at filling this gap by deriving a soft-lubrication perturbation theory, in order to compute all the forces and torque for this problem, at first order in dimensionless compliance. 

The article is organized as follows. First, we introduce the soft-lubrication framework for a sphere translating near a soft planar surface, both in the normal and tangential directions. The substrate deformation is assumed to follow the constitutive response of a linear elastic semi-infinite material. Then, we perform a perturbative approach, assuming the substrate deformation to be small with respect to the fluid-gap thickness, which allows us to find the normal and tangential forces as well as the torque experienced by the sphere, at first order in dimensionless compliance. Finally, we discuss the rotation of the sphere, before providing concluding remarks. Besides, in the appendix, the EHD forces are computed analytically using the Lorentz reciprocal theorem, while the procedure introduced in the main text is reproduced for the compressible and incompressible responses of a thin material.

\section{Model}
\label{sec:model}

\begin{figure}
\centering
\includegraphics[width=0.5\columnwidth]{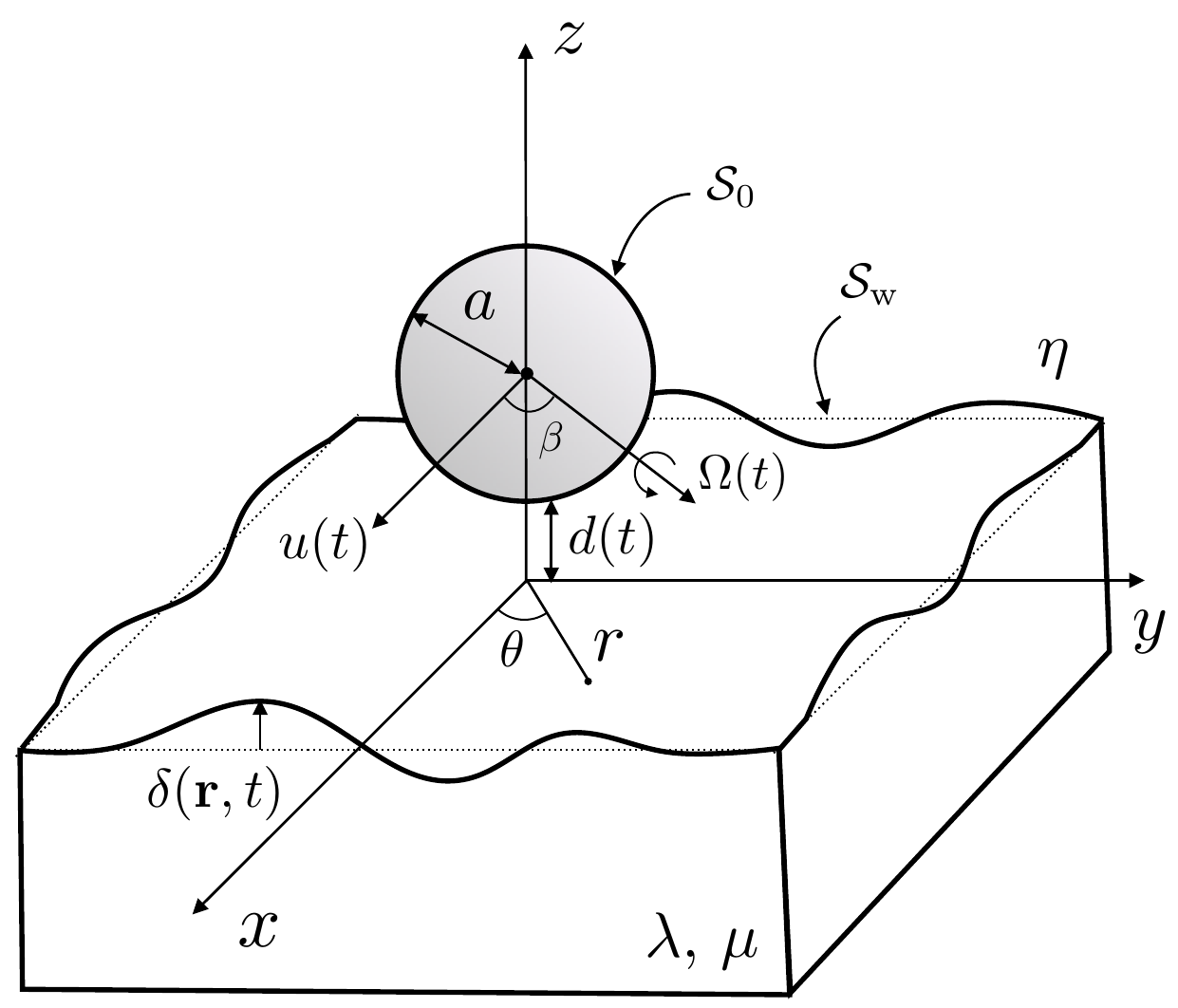}
\caption{Schematic of the system. A rigid sphere of surface $\mathcal S_0$ is freely moving in a viscous fluid, near a soft wall of surface $\mathcal S_{\textrm{w}}$ in the flat undeformed state. The lubrication pressure field deforms the latter which induces an elastohydrodynamic coupling, with forces and torque exerted on the sphere. Note that the surface deformation is magnified for clarity, but that we restrict the analysis to the $\delta \ll d$ case.}
\label{fig:scheme}
\end{figure} 

The system is depicted in Fig.~\ref{fig:scheme}. We consider a sphere of radius $a$, immersed in a Newtonian fluid of dynamic shear viscosity $\eta$ and density $\rho$. The sphere is moving with a tangential velocity $\boldsymbol{u}(t) = u(t) \, \boldsymbol{e}_x$ directed along the $x$-axis (by definition of the latter axis), where $\boldsymbol{e}_j$ denotes the unit vector along $j$. In this first part, we assume that the sphere does not rotate, \textit{i.e.} the angular velocity reads $\boldsymbol{\Omega}=\boldsymbol{0}$. The sphere is placed at a time-dependent distance $d(t)$ (thus a $\dot{d}\, \boldsymbol{e}_z$ normal velocity of the sphere) of an isotropic and homogeneous linear elastic substrate of Lam\'{e} coefficients $\lambda$ and $\mu$, with a reference undeformed flat surface in the $xy$ plan at $z=0$. We suppose that the sphere-wall distance is small with respect to the sphere radius, such that the lubrication approximation is valid. The fluid inertia is neglected here. Specifically, we assume $\textrm{Re}\, (d/a) \ll 1$, with the Reynolds number $\textrm{Re} = \rho u a /\eta$. Furthermore, we suppose that the typical time scale of variation of the sphere velocity is much larger than the diffusion time scale of vorticity that scales as $d^2/(\eta/\rho)$, such that the flow is described by the steady Stokes equations. This amounts to assuming that $\lvert \dot{u}/u \rvert \ll \eta/(\rho d^2)$ and $\lvert \ddot{d}/\dot{d} \rvert \ll \eta/(\rho d^2)$. No-slip boundary conditions are assumed at both the sphere and wall surfaces. Finally, the system is equivalent to a sphere at rest near a wall translating with a $-(d(t)\boldsymbol{e}_z+\boldsymbol{u}(t))$ velocity. In such a framework, the fluid velocity field can be written as: 
\begin{equation}
\label{eq:velocityfields}
\boldsymbol{v}(\boldsymbol{r},z,t) = \frac{\boldsymbol{\nabla} p(\boldsymbol{r},t)}{2\eta} (z - h_0(r,t))(z-\delta(\boldsymbol{r},t)) - \boldsymbol{u}(t)\frac{h_0(r,t)-z}{h_0(r,t)-\delta(\boldsymbol{r},t)},
\end{equation}
where $\boldsymbol{r}=(r,\theta)$ is the position in the tangential plane $xy$, $\boldsymbol{\nabla}$ is the 2D gradient operator on $xy$, $\delta(\boldsymbol{r},t)$ is the substrate deformation, and $z = h_0(r,t)$ is the sphere surface. Near contact, the latter can be approximated by its parabolic expansion $h_0(r,t) \simeq d(t) + r^2/(2a)$. Volume conservation further leads to the Reynolds equation:
\begin{equation}
\label{eq:reynolds}
\partial_t h(\boldsymbol{r}, t) = \boldsymbol{\nabla} \cdot \bigg(\frac{h^3(\boldsymbol{r}, t)}{12\eta} \boldsymbol{\nabla} p(\boldsymbol{r}, t) + \frac{h(\boldsymbol{r}, t)}{2}\boldsymbol{u}(t) \bigg),
\end{equation}
where $h(\boldsymbol{r}, t) = h_0(r,t) - \delta(\boldsymbol{r}, t)$ is the fluid-gap thickness. In this first part, we assume that the constitutive elastic response is linear and instantaneous, and that the substrate is a semi-infinite medium, such that the deformation reads~\cite{davis1986elastohydrodynamic}:
\begin{equation}
\label{eq:elasticity_semiinf}
\delta(\boldsymbol{r}, t) = -\frac{(\lambda+2\mu)}{4\pi \mu(\lambda+\mu)} \int_{\mathbb{R}^2}\textrm{d}^2\boldsymbol{x}  \frac{p(\boldsymbol{x},t)}{\lvert\boldsymbol{r} - \boldsymbol{x} \rvert}.
\end{equation}
We non-dimensionalize the problem through: 
\begin{equation}
\label{eq:non-dimensionalization}
h(\boldsymbol{r}, t) = d^* H(\boldsymbol{R}, T), \quad \boldsymbol{r} = \sqrt{2ad^*} \, \boldsymbol{R}, \quad d(t) = d^* D(T), \quad \delta(\boldsymbol{r}, t) = d^* \Delta(\boldsymbol{R}, T),
\end{equation}
\begin{equation}
\label{eq:non-dimensionalization2}
p(\boldsymbol{r}, t) = \frac{\eta u^* \sqrt{2ad^*}}{d^{*2}}\, P(\boldsymbol{R}, T), \quad \boldsymbol{u}(t) = u^* \, U(T) \, \boldsymbol{e}_x, \quad \boldsymbol{v} = u^* \boldsymbol{V}, \quad t = \frac{\sqrt{2ad^*}}{u^*}\, T.
\end{equation}
where $d^*$ and $u^*$ are characteristic fluid-gap distance and tangential velocity, respectively. The governing equations are then: 
\begin{equation}
\label{eq:DL_Reynold}
12\partial_T H(\boldsymbol{R}, T) = \boldsymbol{\nabla} \cdot \bigg(H^3(\boldsymbol{R}, T) \boldsymbol{\nabla} P(\boldsymbol{R}, T) + 6 H(\boldsymbol{R}, T)\boldsymbol{U}(T) \bigg),
\end{equation}
\begin{equation}
H(\boldsymbol{R}, T) = D(T) + R^2 - \Delta(\boldsymbol{R}, T),
\end{equation}
and:
\begin{equation}
\label{eq:DL_Elasticity}
\Delta(\boldsymbol{R}, T) = -\kappa \int_{\mathbb{R}^2} \textrm{d}^2\boldsymbol{X} \frac{P(\boldsymbol{X},T)}{4\pi \lvert\boldsymbol{R} - \boldsymbol{X} \rvert},
\end{equation}
where we introduced the dimensionless compliance:
\begin{equation}
\kappa=\frac{2\eta u^* a(\lambda+2\mu)}{d^{*2}\mu(\lambda+\mu)}.
\end{equation}
The latter is the only dimensionless parameter in the problem. When $\kappa$ is small with respect to unity, it corresponds to the ratio between two length scales: the typical substrate deformation $\delta \sim \frac{2\eta u^* a(\lambda+2\mu)}{d^*\mu(\lambda+\mu)}$ induced by a tangential velocity $u^*$, and the typical fluid-gap thickness $d^*$. All along the article, we focus on the small-deformation regime of soft-lubrication where $\kappa\ll1$~\citep{essink2021regimes}.

\section{Perturbation theory}
\label{sec:perturbation}
We perform a perturbation analysis at small $\kappa$~\citep{sekimoto1993mechanism,beaucourt2004optimal,skotheim2004soft,skotheim2005soft,urzay2007elastohydrodynamic,urzay2010asymptotic,salez2015elastohydrodynamics,pandey2016lubrication,rallabandi2017rotation,Saintyves2020,zhang2020direct}, as follows:
\begin{equation}
\label{eq:perturbation1}
H(\boldsymbol{R}, T) = H_0(\boldsymbol{R},T) + \kappa \, H_1(\boldsymbol{R}, T) + O(\kappa^2),
\end{equation}
\begin{equation}
\label{eq:perturbation2}
P(\boldsymbol{R}, T) = P_0(\boldsymbol{R},T) + \kappa \, P_1(\boldsymbol{R}, T) + O(\kappa^2),
\end{equation}
where the subscript $0$ corresponds to the solution for a rigid wall, with $H_0(\boldsymbol{R},T) = D(T) + R^2$. 

\subsection{Zeroth-order solution: rigid wall}
\label{sec:leading-order}

\begin{figure}
\centering
\includegraphics[width=0.6\columnwidth]{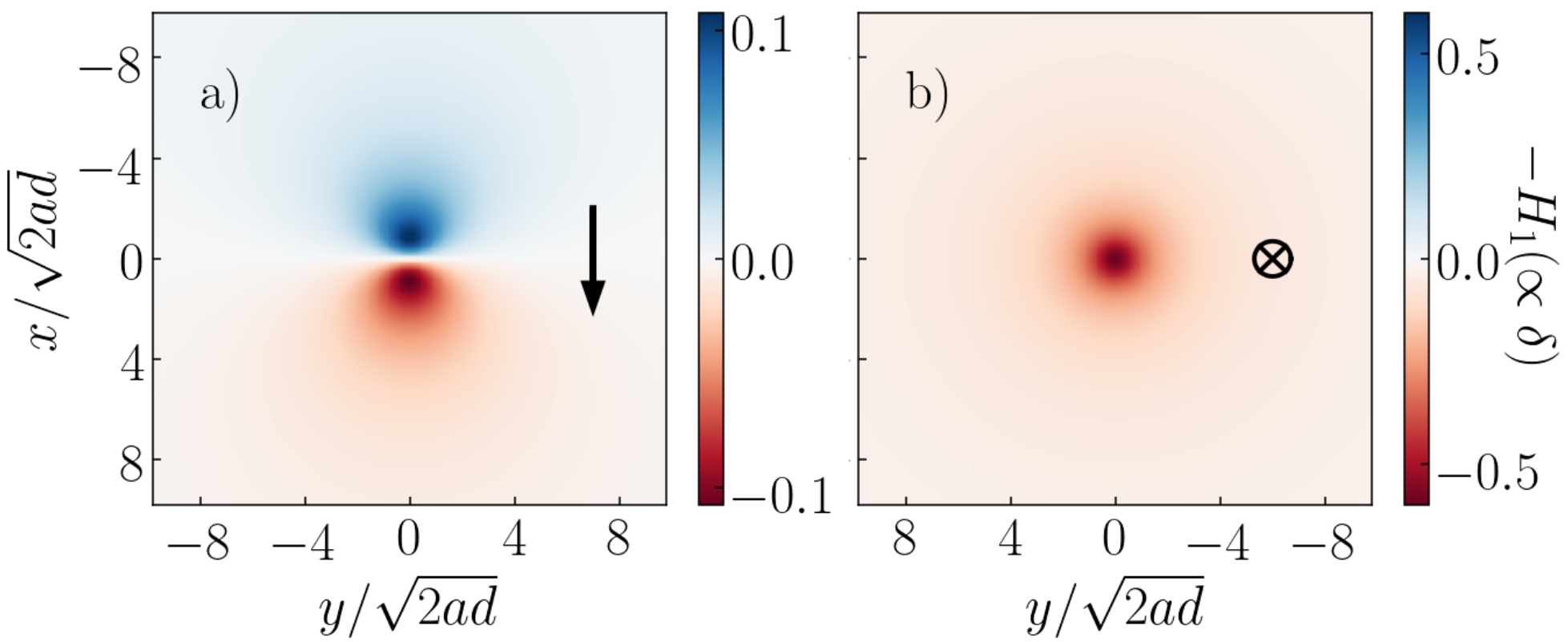}
\caption{ Dimensionless deformation fields at the free surface of the soft substrate, for a sphere placed at a unit distance $D = 1$ and for two modes of motion: a) the sphere velocity is directed tangentially to the substrate, along the $x$-axis (see black arrow), and is fixed to a unit value $U = 1$; b) the sphere is approaching the substrate normally (see black cross) with a unit velocity $\dot{D} = -1$.}
\label{fig:H_1}
\end{figure}

Equation \eqref{eq:DL_Reynold} reads at zeroth order $O(\kappa^0)$:
\begin{equation}
\label{eq:leading-order-pressure-nabla}
12 \dot{D} =\boldsymbol{\nabla} \cdot \bigg(H_0^3 \boldsymbol{\nabla} P_0 + 6 H_0\boldsymbol{U} \bigg).
\end{equation}
In polar coordinates, Eq.~\eqref{eq:leading-order-pressure-nabla} can be rewritten as:
\begin{equation}
\label{eq:leadingorder-pressure}
\mathcal{L}. P_0 = R^2 \partial_R^2 P_0 + \bigg(R + \frac{6R^3}{D+R^2} \bigg) \partial_R P_0 + \partial_\theta^2 P_0 = \frac{R^2}{(D + R^2)^3} \, \bigg( 12 \dot{D} - 12 R \cos\theta \, U \bigg),
\end{equation}
where $\mathcal{L}$ is a linear operator. We solve this equation, using an angular-mode decomposition:
\begin{equation}
P_0(\boldsymbol{R},T) = P_0^{(0)}(R,T) + P_0^{(1)}(R,T) \cos\theta, 
\end{equation}
where the two coefficients are solutions of the ordinary differential equations:
\begin{subequations}
\begin{equation}
R^2 \frac{\textrm{d}^2 P_0^{(0)}}{\textrm{d}R^2}  + \bigg( R + \frac{6R^3}{D+R^2} \bigg) \frac{\textrm{d} P_0^{(0)}}{\textrm{d}R}  = 12 \frac{R^2 \dot{D}}{(D + R^2)^3},
\end{equation}
\end{subequations}
\begin{subequations}
\begin{equation}
R^2 \frac{\textrm{d}^2 P_0^{(1)}}{\textrm{d}R^2} + \bigg( R + \frac{6R^3}{D+R^2} \bigg) \frac{\textrm{d} P_0^{(1)}}{\textrm{d}R} - P_0^{(1)}  = -12 \frac{R^3 U}{(D + R^2)^3}.
\end{equation}
\end{subequations}
In accordance with the boundary conditions, $P(R\rightarrow \infty) = 0$ and $P(R = 0) < \infty$, the solution is thus:
\begin{equation}
\label{eq:leadingorder-pressure_expression}
P_0(\boldsymbol{R},T) = -\frac{3 \dot{D}}{2 (D+R^2)^2} + \frac{6 R U \cos\theta}{5(D+ R^2)^2}.
\end{equation}
The first-order substrate deformation $H_1$ can then be computed from Eq.~\eqref{eq:DL_Elasticity} at order $O(\kappa)$:
\begin{equation}
H_1(\boldsymbol{R}, T) = \int_{\mathbb{R}^2} \textrm{d}^2\boldsymbol{X} \frac{P_0(\boldsymbol{X},T)}{4\pi \lvert\boldsymbol{R} - \boldsymbol{X} \rvert}.
\end{equation}
Using \textit{e.g.} the spatial Fourier transform $\tilde{H}_1(\boldsymbol{K}) = \int_{\mathbb{R}^2} H_1(\boldsymbol{R})e^{-i\boldsymbol{R}\cdot\boldsymbol{K}} \textrm{d}^2\boldsymbol{R}$, we find:
\begin{equation}
\label{eq:leading_order_deformation}
H_1(\boldsymbol{R}, T) = -\frac{3\dot{D}}{8\sqrt{D}} \, \frac{\mathcal{E}(-R^2/D)}{D+ R^2} + \frac{3U}{10R\sqrt{D}} \, \bigg(- \frac{D\, \mathcal{E}(-R^2/D)}{D+R^2} + \mathcal{K}(-R^2/D) \bigg) \cos \theta, 
\end{equation}
where $\mathcal{K}$ and $\mathcal{E}$ are the complete elliptic integrals of the first and second kinds~\citep{abramowitz1964handbook}. The dimensionless substrate deformations are plotted in Fig.~\ref{fig:H_1}. In Fig.~\ref{fig:H_1}a), the sphere is moving tangentially to the substrate with a unit velocity $U=1$. The deformation exhibits a dipolar symmetry, with a negative sign (\textit{i.e.} the substrate is compressed) at the front. Besides, the isotropic term generated by a sphere moving normally to the substrate is shown in Fig.~\ref{fig:H_1}b). In particular, for a sphere approaching the substrate, the latter is compressed.

\subsection{First-order solution}
\begin{figure}
\centering
\includegraphics[width=0.9\columnwidth]{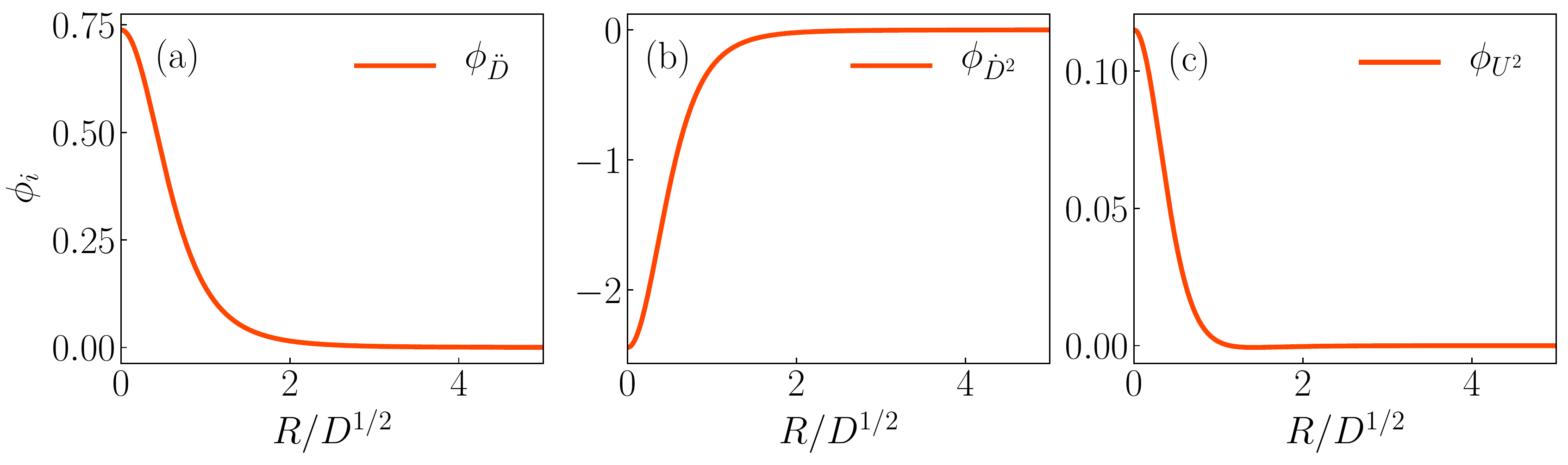}
\caption{Scaling functions for $P_1^{(0)}$ (see Eq.~\eqref{eq:AuxP_1^0}), obtained from numerical integration of Eq.~\eqref{eq:P_1^0}, with the boundary conditions $\partial_R P_1^{(0)} (R = 0,T) = 0$ and $P_1^{(0)}(R \rightarrow \infty,T)=0$.}
\label{fig:P_1^0}
\end{figure}

\begin{figure}
\centering
\includegraphics[width=0.6\columnwidth]{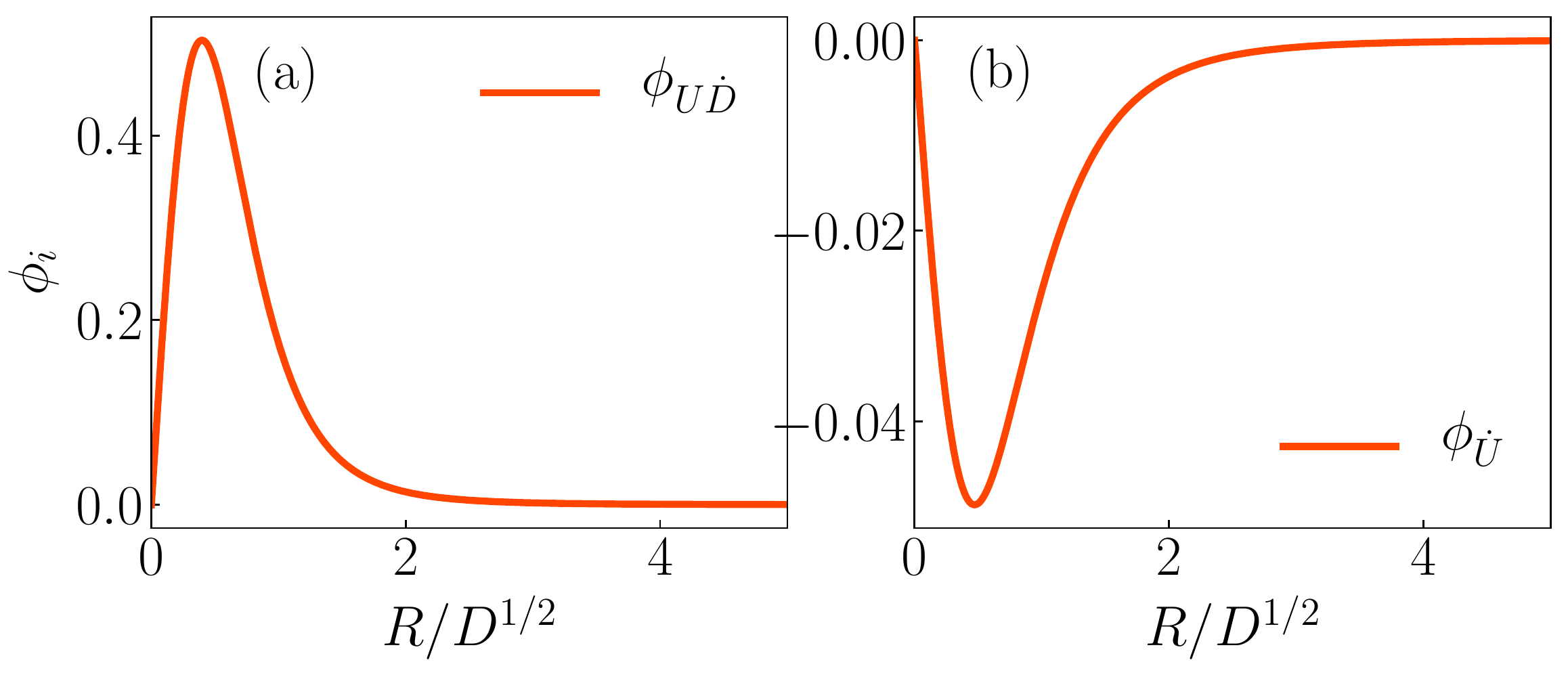}
\caption{Scaling functions for $P_1^{(1)}$ (see Eq.~\eqref{eq:AuxP_1^1}), obtained from numerical integration of Eq.~\eqref{eq:P_1^1}, with the boundary conditions $P_1^{(1)} (R = 0,T) = 0 $ and  $P_1^{(1)} (R \rightarrow \infty,T) =0$.}
\label{fig:P_1^1}
\end{figure}

We can now compute the first-order pressure field $P_1$, from Eq.~\eqref{eq:DL_Reynold} at order $O(\kappa)$:
\begin{equation}
\label{eq:nextleadingorderpresssure}
12\partial_T H_1 = \boldsymbol{\nabla} \cdot \bigg(H_0^3 \boldsymbol{\nabla} P_1 + 3H_0^2 H_1 \boldsymbol{\nabla} P_0 + 6 H_1\boldsymbol{U} \bigg). 
\end{equation}
Invoking the same linear operator $\mathcal{L}$ as in Eq.~\eqref{eq:leadingorder-pressure}, we can rewrite Eq.~\eqref{eq:nextleadingorderpresssure} as:
\begin{equation}
\label{eq:nextleadpres-polar}
\mathcal{L}.P_1 = \frac{R^2}{H_0^3} \bigg(12 \partial_T H_1 - \boldsymbol{\nabla} \cdot \bigg[ 3H_0^2 H_1 \boldsymbol{\nabla} P_0 + 6 H_1\boldsymbol{U}\bigg] \bigg).
\end{equation}
We then expand all the terms in the right-hand side of Eq.~\eqref{eq:nextleadpres-polar}, and we perform once again the angular-mode decomposition:
\begin{equation}
\mathcal{L}.P_1 = F_0(R,T) + F_1(R,T) \cos\theta + F_2(R,T) \cos2\theta,
\end{equation}
where we have introduced the auxiliary functions: 
\begin{equation}
\begin{split}
F_0(R,T) &= \frac{18R^2 U^2}{25 D^{1/2}(D+R^2)^6}  \bigg[ (-10D^2 +2DR^2) \, \mathcal{E}\left(-\frac{R^2}{D}\right)  + (8D^2+7DR^2-R^4) \, \mathcal{K}\left(-\frac{R^2}{D}\right)   \bigg] \\
& \, \, + \frac{9R^2 \dot{D}^2}{4 D^{3/2}(D+R^2)^6} \bigg[ (13D^2 + 3R^2D + 2R^4)\, \mathcal{E}\left(-\frac{R^2}{D}\right)  + (-4D^2 - 5R^2D-R^4) \, \mathcal{K}\left(-\frac{R^2}{D}\right) \bigg] \\
& \, \,  -\frac{9R^2 \ddot{D}\, \mathcal{E}\left(-\frac{R^2}{D}\right)}{2 D^{1/2}(D+R^2)^4},
\end{split}
\end{equation}
and:
\begin{equation}
\begin{split}
\label{eq:F_1}
F_1(R,T) &= -\frac{27 R U\dot{D} }{5 D^{1/2}(D+R^2)^6}  \bigg[ (-2D^2 +7DR^2+R^4) \, \mathcal{E}\left(-\frac{R^2}{D}\right)  + 2(D+R^2)(D-R^2) \, \mathcal{K}\left(-\frac{R^2}{D}\right)   \bigg] \\
& \, \, - \frac{18R \dot{U}}{5 D^{1/2}(D+R^2)^4} \bigg[- D\, \mathcal{E}\left(-\frac{R^2}{D}\right)  + (D + R^2) \, \mathcal{K}\left(-\frac{R^2}{D}\right) \bigg].
\end{split}
\end{equation}
We note that we have not provided $F_2$ as it does not contribute in the forces and torque. We also note that, by setting $D(T) = 1$ in the latter expressions, we self-consistently recover the expression of~\cite{zhang2020direct}. Invoking the angular-mode decomposition $P_1(\boldsymbol{R},T) = P_1^{(0)}(R,T) + P_1^{(1)}(R,T)\cos \theta + P_1^{(2)}(R,T) \cos 2\theta$, we get in particular:
\begin{equation}
\label{eq:P_1^0}
R^2 \frac{\textrm{d}^2 P_1^{(0)}}{\textrm{d}R^2}  + \bigg( R + \frac{6R^3}{D+R^2} \bigg) \frac{\textrm{d} P_1^{(0)}}{\textrm{d}R}  = F_0(R,T),
\end{equation}
\begin{equation}
\label{eq:P_1^1}
R^2 \frac{\textrm{d}^2 P_1^{(1)}}{\textrm{d}R^2}  + \bigg( R + \frac{6R^3}{D+R^2} \bigg) \frac{\textrm{d} P_1^{(1)}}{\textrm{d}R} - P_1^{(1)}  = F_1(R,T).
\end{equation}
Using scaling arguments, we can write the two relevant first-order pressure components $P_1^{(i)}$ as: 
\begin{equation}
\label{eq:AuxP_1^0}
P_1^{(0)} = \frac{U^2}{D^{7/2}} \phi_{U^2}(R/\sqrt{D}) + \frac{\dot{D}^2}{D^{9/2}} \phi_{\dot{D}^2}(R/\sqrt{D}) + \frac{\ddot{D}}{D^{7/2}} \phi_{\ddot{D}}(R/\sqrt{D}),
\end{equation}
and:
\begin{equation}
\label{eq:AuxP_1^1}
P_1^{(1)} = \frac{U\dot{D} }{D^4} \phi_{U \dot{D}}(R/\sqrt{D}) + \frac{\dot{U}}{D^3} \phi_{\dot{U}}(R/\sqrt{D}),
\end{equation}
where the $\phi_i$ are five dimensionless scaling functions that depend on the self-similar variable $R/\sqrt{D}$ only. Equations~\eqref{eq:AuxP_1^0} and \eqref{eq:AuxP_1^1} can be solved numerically with a Runge-Kutta algorithm, and a shooting parameter in order to ensure the boundary condition $P_1(R \rightarrow \infty, \theta, T) = 0$. All the scaling functions are plotted in Figs.~\ref{fig:P_1^0}~and~\ref{fig:P_1^1}. 

As a remark, we recall that the substrate deformation is induced by the flow, and that at first order it is linear in the velocity field (see Eq.~\eqref{eq:leading_order_deformation}). Moreover, the volume-conservation equation involves the time derivative of the fluid-layer thickness, and thus in particular the time derivative of the substrate deformation. As a consequence, when calculating the first-order EHD pressure field, we find terms (and thus forces and torques) that are proportional to the acceleration components $\ddot{D}$ and $\dot{U}$ of the sphere. At first sight, these original inertial-like features
might seem inconsistent with steady Stokes flows, but are in fact independent of the fluid density and solely induced by the intimate EHD coupling. 

\subsection{Forces and torque}
\label{sec:force}
The force $\boldsymbol{F}$ exerted by the fluid on the sphere is given by:
\begin{equation}
\label{eq:force}
\boldsymbol{F}=  \int_{\mathcal{S}_0}  \boldsymbol{n} \cdot \boldsymbol{\sigma} \, \mathrm{d}s,
\end{equation}
where $\boldsymbol{\sigma} = - p \mathbf{I} + \eta (\boldsymbol{\nabla} \boldsymbol{v} + \boldsymbol{\nabla}  \boldsymbol{v}^\text{T})$ is the fluid stress tensor, $\boldsymbol{n}$ is the unit vector normal to the sphere surface and pointing towards the fluid, and $\mathbf{I}$ is the identity tensor. Within the lubrication approximation, the fluid stress tensor reads $\boldsymbol{\sigma} \simeq - p \mathbf{I} + \eta \boldsymbol{e}_z \partial_z \boldsymbol{v} $. One can then evaluate the normal force, as:
\begin{equation}
\label{eq:normal_force}
\begin{split}
F_z = \int_{\mathbb{R}^2}   p(\boldsymbol{r}) \, \textrm{d}^2 \boldsymbol{r}
= &-\frac{6\pi \eta a^2 \dot{d}}{d} + 0.41623 \frac{\eta^2 u^2(\lambda+2\mu)}{\mu(\lambda+\mu)} \bigg( \frac{a}{d}\bigg)^{5/2} \\ 
&- 41.912 \frac{\eta^2 \dot{d}^2(\lambda+2\mu)}{\mu(\lambda+\mu)} \bigg( \frac{a}{d}\bigg)^{7/2} + 18.499  \frac{\eta^2 \ddot{d} a(\lambda+2\mu)}{\mu(\lambda+\mu)} \bigg( \frac{a}{d}\bigg)^{5/2},
\end{split}
\end{equation}
where the prefactors have been numerically estimated using Eq.~\eqref{eq:AuxP_1^0}. We recover in particular the classical Reynolds force $-6\pi \eta a^2 \dot{d}/d$ at zeroth order, \textit{i.e.} near a rigid wall. We stress that tangential motions do not induce any normal force at zeroth order in $\kappa$, as the corresponding pressure field is antisymmetric in $x$ (see Eq.~\eqref{eq:leadingorder-pressure}). In contrast, such motions do induce a lift force at first order in $\kappa$, due to the symmetry breaking of the contact geometry associated with the elastic deformation. Interestingly as well, normal motions generate a viscous adhesive force at first order in compliance~\citep{wang2020dynamic}. Besides, an original EHD force proportional to the sphere's normal acceleration is also found, as discussed previously. Finally, in the appendix, and following previous works~\citep{rallabandi2017rotation,rallabandi2018membrane,daddi2018reciprocal,masoud2019reciprocal}, we use the Lorentz reciprocal theorem in order to recover the prefactors of Eq.~\eqref{eq:normal_force} analytically, which gives respectively: $\frac{243 \pi ^3}{12800 \sqrt{2}} \approx0.41623$, $\frac{3915 \pi ^3}{2048 \sqrt{2}} \approx 41.912$ and $\frac{27\pi^3}{32\sqrt{2}} \approx 18.499$. We note that the latter is in agreement with the result of the linear-response theory derived in~\cite{leroy2011hydrodynamic}. Furthermore, we recover the lift prefactor (0.416) obtained previously numerically~\citep{zhang2020direct}, as well as analytically in a recently-published work~\citep{kargar2021lift}.

Similarly, the tangential force reads:
\begin{equation}
\label{eq:lateral_force}
\boldsymbol{F}_\parallel = \int_{\mathbb{R}^2}  \, \bigg(-p(\boldsymbol{r},t)\frac{\boldsymbol{r}}{a} - \eta \partial_z \boldsymbol{v} \bigg)_{z = h_0(r,t)}\textrm{d}^2\boldsymbol{r}.
\end{equation}
Using symmetry arguments, we can show that the tangential force is directed along $x$, \textit{i.e.} $\boldsymbol{F}_\parallel =F_x \, \boldsymbol{e}_x$. At small $\kappa$, we further expand it as $F_x \simeq F_{x,0} + \kappa F_{x,1}$, where $F_{x,0}$ is the viscous drag force applied on a sphere near a rigid plane wall, and $\kappa F_{x,1}$ is the first-order EHD correction. The zeroth-order term cannot be evaluated using the lubrication model introduced in the previous section, because the integral in Eq.~\eqref{eq:lateral_force} diverges, as the shear term $\eta \partial_z \boldsymbol{v}$ scales as $\sim r^{-2}$ at large $r$. An exact calculation has been performed using bispherical coordinates and provides a solution in the form of a series expansion~\citep{o1964slow}. Asymptotic-matching methods have also been employed in order to get the asymptotic behavior at small $d/a$~\citep{o1967slow,goldman1967slow}, which reads $F_{x,0} \approx 6 \pi \eta a u \bigg(\frac{8}{15}\log\left(\frac{d}{a}\right) - 0.95429 \bigg)$ (see \citep{chaoui2003creeping} for a high-precision expansion). We note that the sphere's normal velocity does not contribute to the zeroth-order tangential force, as expected by symmetry. 

The first-order EHD correction can be computed with the present model, as the correction pressure field and shear stress scale as $\sim r^{-5}$, at large $r$. It reads:
\begin{equation}
\label{Eq:F_1,x}
\begin{split}
F_{x,1} = 2\pi \eta u^* a   \int_0^\infty   \bigg[& -2  R P_1^{(1)}  - \frac{H_0}{2} \bigg(\partial_R P_1^{(1)}  + \frac{P_1^{(1)}}{R}  \bigg)  \\
&  - \frac{H_1^{(1)}}{2}\partial_R P_0^{(0)} - \frac{H_1^{(0)}}{2} \bigg(\partial_R P_0^{(1)} + \frac{P_0^{(1)}}{R}  \bigg) +2 \frac{U H_1^{(0)}}{H_0^2} \bigg] R \, \textrm{d}R,
\end{split}
\end{equation}
where $H_1^{(i)}$ is the amplitude of the $i^{\textrm{th}}$ mode in the angular-mode decomposition of $H_1$. Evaluating the latter integral numerically, we find:
\begin{equation}
\label{eq:EHD_lateral}
F_x \approx  6 \pi \eta a u \bigg(\frac{8}{15}\log\left(\frac{d}{a}\right) - 0.95429 \bigg) -10.884 \frac{\eta^2 u \dot{d}(\lambda+2\mu)}{\mu(\lambda+\mu)} \bigg( \frac{a}{d}\bigg)^{5/2} + 0.98661 \frac{\eta^2 \dot{u}a(\lambda+2\mu)}{\mu(\lambda+\mu)}\bigg( \frac{a}{d}\bigg)^{3/2}.
\end{equation}
In the appendix, we use again the Lorentz reciprocal theorem in order to compute the first-order EHD force, and we obtain the following analytical expressions for the coefficients of Eq.~\eqref{eq:EHD_lateral}: $ -\frac{3177 \pi ^3}{6400 \sqrt{2}} \simeq -10.884$ and $\frac{9 \pi ^3}{200 \sqrt{2}} \simeq 0.98661$, respectively. 

The torque exerted by the fluid on the sphere, with respect to its center of mass, is given by: 
\begin{equation}
\boldsymbol{T}=  \int_{\mathcal{S}_0}  a\boldsymbol{n} \times (\boldsymbol{n} \cdot \boldsymbol{\sigma}) \, \mathrm{d}s.
\end{equation} 
The latter is directed along the $y$ direction for symmetry reasons, \textit{i.e.} $\boldsymbol{T} =T_y \, \boldsymbol{e}_y$. At small $\kappa$, we further expand it as $T_{y} \simeq T_{y,0} + \kappa T_{y,1}$. For the same reason as with the the viscous drag force near a rigid wall, the viscous torque near a rigid wall cannot be computed within the lubrication model. Using asymptotic-matching methods~\citep{o1967slow,chaoui2003creeping}, it is found to be $T_{y,0} \approx 8\pi \eta u a^2 \bigg(-\frac{1}{10}\log\left(\frac{d}{a}\right) - 0.19296 \bigg)$. In contrast, the first-order EHD correction can be computed with the present model, and reads:
\begin{equation}
\label{Eq:T_1,y}
T_{y,1} = -2\eta u^* a^2 \pi  \int_0^\infty \bigg[   \frac{H_0}{2} \bigg(\partial_R P_1^{(1)}  + \frac{P_1^{(1)}}{R}  \bigg) + \frac{H_1^{(1)}}{2}\partial_R P_0^{(0)} + \frac{H_1^{(0)}}{2} \bigg(\partial_R P_0^{(1)} + \frac{P_0^{(1)}}{R}  \bigg) +2 \frac{U H_1^{(0)}}{H_0^2} \bigg]R \, \textrm{d}R.
\end{equation} 
Evaluating the latter integral numerically, we find:
\begin{equation}
T_y \approx 8\pi \eta u a^2 \bigg(-\frac{1}{10}\log\left(\frac{d}{a}\right) - 0.19296 \bigg) + 10.884 \frac{\eta^2 u a\dot{d}(\lambda+2\mu)}{\mu(\lambda+\mu)} \bigg( \frac{a}{d}\bigg)^{5/2} -0.98661\frac{\eta^2 \dot{u}a^2(\lambda+2\mu)}{\mu(\lambda+\mu)}\bigg( \frac{a}{d}\bigg)^{3/2}.
\end{equation}

So far, we focused on the particular case of a semi-infinite elastic material. In appendices~\ref{app:winkler} and \ref{app:thin-incompressible}, we apply the same soft-lubrication approach to other elastic models describing thin substrates, which are also widespread in practice. We find similar expressions, but with different numerical prefactors and scalings with the sphere-wall distance. 

\section{Rotation}
\label{sec:rotation}
We now add the rotation of the sphere, with angular velocity $\boldsymbol{\Omega}(t)$ in the $xy$ plane (see Fig.~\ref{fig:scheme}), to the previous translational motion. We define $\beta$ as the angle between $\boldsymbol{\Omega}$ and the $x$-axis. We stress that $\boldsymbol{\Omega}$ is not necessarily orthogonal  (\textit{i.e.} $\beta = \pi/2$) to the translation velocity. We discard the rotation along the $z$-axis (\textit{e.g.} for a spinner), because it does not induce any soft-lubrication coupling. Finally, the system is equivalent to a purely rotating sphere with angular velocity $\boldsymbol{\Omega}(t)$, near a wall translating with a $-\boldsymbol{u}(t)$ velocity. In such a framework, the fluid velocity field at the sphere surface is $\boldsymbol{v} = -\boldsymbol{\Omega} \times a\boldsymbol{n}$, and thus $\boldsymbol{v} \simeq -\boldsymbol{\Omega} \times a\boldsymbol{e}_z$. All together, the fluid velocity field is modified as:
\begin{equation}
\boldsymbol{v}(\boldsymbol{r},z,t) = \frac{\boldsymbol{\nabla} p(\boldsymbol{r},t)}{2\eta} (z - h_0(r,t))(z-\delta(\boldsymbol{r},t)) - \boldsymbol{u}(t)\frac{h_0(r,t)-z}{h_0(r,t)-\delta(\boldsymbol{r},t)} + a\boldsymbol{\Omega}(t)\times \boldsymbol{e}_z \, \frac{z-\delta(\boldsymbol{r},t)}{h_0(r,t)-\delta(\boldsymbol{r},t)},
\end{equation}
and the Reynolds equation becomes:
\begin{equation}
\partial_t h(\boldsymbol{r}, t) = \boldsymbol{\nabla} \cdot \bigg(\frac{h^3(\boldsymbol{r}, t)}{12\eta} \boldsymbol{\nabla} p(\boldsymbol{r}, t) + \frac{h(\boldsymbol{r}, t)}{2}\bigg[\underbrace{\boldsymbol{u}(t) -a\boldsymbol{\Omega}(t)\times \boldsymbol{e}_z}_{\tilde{\boldsymbol{u}}}\bigg] \bigg).
\end{equation}
The problem is thus formally equivalent to the one of a sphere that is purely translating with effective velocity $\boldsymbol{\tilde{u}} (t)= \boldsymbol{u}(t) -a\boldsymbol{\Omega}(t)\times  \boldsymbol{e}_z$. Therefore, we can directly apply the results from the previous sections, and write all the forces and torque exerted on the sphere, as:
\begin{equation}
\label{eq:normal_force_EHD_rotation}
F_z = -\frac{6\pi\eta a^2\dot{d}}{d} + \frac{243 \pi ^3}{12800 \sqrt{2}} \frac{\eta^2 \lvert \boldsymbol{u} -a\boldsymbol{\Omega}\times \boldsymbol{e}_z \rvert^2}{\mu} \bigg( \frac{a}{d}\bigg)^{5/2} - \frac{3915 \pi ^3}{2048 \sqrt{2}} \frac{\eta^2 \dot{d}^2}{\mu} \bigg( \frac{a}{d}\bigg)^{7/2} + \frac{27\pi^3}{32\sqrt{2}}  \frac{\eta^2 \ddot{d} a}{\mu} \bigg( \frac{a}{d}\bigg)^{5/2},
\end{equation}
\label{eq:lateral_force_EHD_rotation}
\begin{equation}
\begin{split}
\boldsymbol{F}_\parallel &=  6 \pi \eta a \boldsymbol{u} \bigg[\frac{8}{15}\log \bigg(\frac{d}{a} \bigg) - 0.95429 \bigg] +6\pi\eta a^2 \boldsymbol{e}_z \times \boldsymbol{\Omega} \bigg[ \frac{2}{15} \log \bigg( \frac{d}{a} \bigg) + 0.25725 \bigg] \\
&-\frac{3177 \pi ^3}{6400 \sqrt{2}} \frac{\eta^2 (\boldsymbol{u} -a\boldsymbol{\Omega}\times \boldsymbol{e}_z) \dot{d}}{\mu} \bigg( \frac{a}{d}\bigg)^{5/2} +\frac{9 \pi ^3}{200 \sqrt{2}} \frac{\eta^2 (\boldsymbol{\dot{u}} -a\boldsymbol{\dot{\Omega}} \times \boldsymbol{e}_z) a}{\mu}\bigg( \frac{a}{d}\bigg)^{3/2},
\end{split}
\end{equation}
and:
\begin{equation}
\label{eq:torque_EHD_rotation}
\begin{split}
\boldsymbol{T}_\parallel &=  8  \pi \eta a^2 \boldsymbol{e}_z\times \boldsymbol{u} \bigg[-\frac{1}{10}\log \bigg(\frac{d}{a} \bigg) - 0.19296 \bigg] +8\pi\eta a^3 \boldsymbol{\Omega} \bigg[ \frac{2}{5} \log \bigg( \frac{d}{a} \bigg) - 0.37085\bigg]\\
&+\frac{3177 \pi ^3}{6400 \sqrt{2}} \frac{\eta^2 (\boldsymbol{u} -a\boldsymbol{\Omega} \times \boldsymbol{e}_z) a\dot{d}}{\mu} \bigg( \frac{a}{d}\bigg)^{5/2} -\frac{9 \pi ^3}{200 \sqrt{2}} \frac{\eta^2(\boldsymbol{\dot{u}} -a\boldsymbol{\dot{\Omega}}\times \boldsymbol{e}_z)a^2}{\mu}\bigg( \frac{a}{d}\bigg)^{3/2},
\end{split}
\end{equation}
where we have invoked the force and torque induced by the rotation of a sphere near a rigid wall~\citep{goldman1967slow,urzay2010asymptotic} and where the analytical prefactors are computed in the appendix. We stress that the expressions of the EHD forces and torque for a sphere purely translating near thin elastic substrates, as derived in appendices~\ref{app:winkler} and \ref{app:thin-incompressible}, can be generalized to further include the sphere's rotation by similarly following the transformation $\boldsymbol{u}(t) \rightarrow \boldsymbol{u}(t) -a\boldsymbol{\Omega}(t)\times  \boldsymbol{e}_z$.

\section{Conclusion} 
\label{sec:conclusion}
 
We developed a soft-lubrication model in order to compute the EHD interactions exerted on an immersed sphere undergoing both translational and rotational motions near various types of elastic walls. The deformation of the surface was assumed to be small, which allowed us to employ a perturbation analysis in order to obtain the leading-order EHD forces and torque. The obtained interaction matrix exhibits a qualitatively similar form as the one found for a two-dimensional cylinder moving near a thin compressible substrate~\citep{salez2015elastohydrodynamics}. In both cases, the EHD coupling is nonlinear and generates quadratic terms in the sphere velocity, thus breaking the time-reversal symmetry of the Stokes equations. In addition, original inertial-like terms proportional to the acceleration of the sphere are found -- despite the assumption of steady flows. Therefore, while the quantitative details such as numerical prefactors and exponents differ in 3D and when using more realistic constitutive elastic responses, we expect that the typical zoology of trajectories identified previously~\citep{salez2015elastohydrodynamics} will also hold for spherical objects -- and will even be extended with the added degree of freedom. As such, the asymptotic predictions obtained here may open new perspectives in colloidal science and biophysics, through the understanding and control of the emerging interactions within soft confinement or assemblies.

\section*{Acknowledgments}
We thank Abelhamid Maali, Zaicheng Zhang, Howard Stone and Bhargav Rallabandi for stimulating discussions. 

\section*{Fundings}
The authors acknowledge funding from the Agence Nationale de la Recherche (ANR-21-ERCC-0010-01 \textit{EMetBrown}) and from the Jean Langlois foundation.

\section*{Declaration of Interests} The authors report no conflict of interest.

\appendix 
\section{Lorentz reciprocal theorem: normal force}
\label{app:Lorentz-Normal}
In this appendix, we compute the first-order normal EHD force using the Lorentz reciprocal theorem for Stokes flows~\citep{rallabandi2017rotation,rallabandi2018membrane,daddi2018reciprocal,masoud2019reciprocal}, in order to recover analytically the numerical prefactors obtained in the main text. To do so, we introduce the model problem of a sphere moving in a viscous fluid and towards an immobile, rigid, planar surface. We note $\boldsymbol{\hat{V}}_\perp = -\hat{V}_\perp\boldsymbol{e}_z$ the velocity at the particle surface $\mathcal{S}_0$, and we assume a no-slip boundary condition at the undeformed wall surface $\mathcal{S}_\text{w}$ located at $z = 0$ (see Fig.~\ref{fig:scheme}). The viscous stress and velocity fields of the model problem follow the steady,  incompressible Stokes equations $\boldsymbol{\nabla} \cdot \boldsymbol{\hat{\sigma}}_\perp = \boldsymbol{0}$ and $\boldsymbol{\nabla} \cdot \boldsymbol{\hat{v}}_\perp =0$, and we use the lubrication approximation. In this framework, the stress tensor is $\boldsymbol{\hat{\sigma}}_\perp \simeq -\hat{p}_\perp \mathbf{I} + \eta \boldsymbol{e}_z \partial_z \boldsymbol{\hat{v}}_\perp$, with:
\begin{equation}
\label{eq:model-solution}
\hat{p}_\perp(\boldsymbol{r}) = \frac{3\eta \hat{V}_\perp a}{\hat{h}^2(\boldsymbol{r})}, \quad \quad \boldsymbol{\hat{v}}_\perp(\boldsymbol{r},z) = \frac{\boldsymbol{\nabla}\hat{p}_\perp(\boldsymbol{r})}{2\eta} z(z-\hat{h}(\boldsymbol{r})), \quad \quad \hat{h}(\boldsymbol{r}) = d + \frac{\boldsymbol{r}^2}{2a}.
\end{equation}
The Lorentz reciprocal theorem states that:
\begin{equation}
\label{eq:Lorentz}
\int_\mathcal{S}   \boldsymbol{n} \cdot \boldsymbol{\sigma} \cdot \boldsymbol{\hat{v}}_\perp \, \mathrm{d}s = \int_\mathcal{S}   \boldsymbol{n} \cdot \boldsymbol{\hat{\sigma}}_\perp \cdot \boldsymbol{v} \, \mathrm{d}s, 
\end{equation}
where $\mathcal{S} = \mathcal{S}_0 + \mathcal{S}_\text{w} + \mathcal{S}_\infty$ is the total surface bounding the flow, and $\mathcal{S}_\infty$ is the surface located at $\boldsymbol{r} \rightarrow \infty$. The latter does not contribute here. Using the boundary conditions for the model problem, we get:
\begin{equation}
\label{eq:Lorentz_z-projected}
\boldsymbol{\hat{V}}_\perp\cdot \boldsymbol{F} = -\hat{V}_\perp F_z = \int_{\mathcal{S}}   \boldsymbol{n} \cdot \boldsymbol{\hat{\sigma}}_\perp \cdot \boldsymbol{v} \, \mathrm{d}s.
\end{equation}
To find the force exerted on the sphere in the real problem, one needs to specify the boundary conditions for the real velocity field. Here, we assume that the sphere does not rotate, and we describe the flow in the translating reference frame of the particle. The no-slip boundary condition thus reads $\boldsymbol{v} = \boldsymbol{0}$ on $\mathcal{S}_0$. We further assume a small deformation of the wall, so that the velocity field at the undeformed wall surface can be obtained using the Taylor expansion:
\begin{equation}
\begin{split}
\boldsymbol{v}\vert_{z = 0} &= \boldsymbol{v}\vert_{z = \delta} - \delta \partial_z \boldsymbol{v}_0\vert_{z = 0} \\
& = -u\boldsymbol{e}_x - \dot{d}\boldsymbol{e}_z + (\partial_t  - u\partial_x ) \delta \boldsymbol{e}_z- \delta \partial_z \boldsymbol{v}_0\vert_{z = 0},
\end{split}
\end{equation}
where $\boldsymbol{v}_0$ is the zeroth-order velocity field near a rigid surface. Using results from the main text, we find:
\begin{equation}
\partial_z \boldsymbol{v}_0\vert_{z = 0} = -\frac{3\dot{d}r}{(d+\frac{r^2}{2a})^2} \boldsymbol{e}_r + \frac{2u}{5(d+\frac{r^2}{2a})} \left(\left(7-\frac{6d}{d+\frac{r^2}{2a}}\right)\cos\theta \boldsymbol{e}_r - \sin\theta\boldsymbol{e}_\theta \right).
\end{equation}
Combining Eqs.~\eqref{eq:model-solution}~and~\eqref{eq:Lorentz_z-projected}, we find the normal force:
\begin{equation}
F_z = \frac{1}{\hat{V}_\perp} \int_{\mathbb{R}^2} \bigg(\hat{p}_\perp (-\dot{d} + \partial_t \delta - u\partial_x \delta)+\eta \partial_z \boldsymbol{\hat{v}}_\perp\vert_{z = 0}\cdot \partial_z \boldsymbol{v}_0  \vert_{z = 0} \delta \bigg)\, \textrm{d}\boldsymbol{r}.
\end{equation}
After some algebra, and computing the integral in Fourier space, we recover the same expression as in Eq.~\eqref{eq:normal_force}, that reads:
\begin{equation}
F_z = -\frac{6\pi \eta a^2 \dot{d}}{d} + A \frac{\eta^2 u^2(\lambda+2\mu)}{\mu(\lambda+\mu)} \bigg( \frac{a}{d}\bigg)^{5/2} - B \frac{\eta^2 \dot{d}^2(\lambda+2\mu)}{\mu(\lambda+\mu)} \bigg( \frac{a}{d}\bigg)^{7/2} + C  \frac{\eta^2 \ddot{d} a(\lambda+2\mu)}{\mu(\lambda+\mu)} \bigg( \frac{a}{d}\bigg)^{5/2},
\end{equation}
where the numerical coefficients $A, B, C$ can be found analytically as:
\begin{equation}
A = \frac{9\pi}{25\sqrt{2}} \int_0^\infty k^2 K_0(k)\bigg(-2K_1(k) + kK_2(k) \bigg) \,k \textrm{d}k = \frac{243 \pi ^3}{12800 \sqrt{2}},
\end{equation}
\begin{equation}
B =9\pi\sqrt{2} \int_0^\infty k^2 K_1(k)\bigg(K_2(k)-\frac{kK_3(k)}{8}\bigg)  \,k \textrm{d}k = \frac{3915 \pi ^3}{2048 \sqrt{2}},
\end{equation}
\begin{equation}
C = \frac{9\sqrt{2}\pi}{2} \int_0^\infty k^2 K_1^2(k) \, \textrm{d}k = \frac{27\pi^3}{32\sqrt{2}},
\end{equation}
and where $K_i$ is the modified Bessel function of the second kind of order $i$~\citep{abramowitz1964handbook}.

\section{Lorentz reciprocal theorem: tangential force}
\label{app:Lorentz-Tangent}

In order to compute the tangential force acting on the particle, we apply the Lorentz reciprocal theorem, but we introduce a different model problem with respect to the previous section. We consider a sphere translating parallel to a rigid immobile substrate, with a velocity $\hat{V}_\parallel$ along the $x$-axis, and no-slip boundary conditions at both the sphere and substrate surfaces. The velocity and stress fields are denoted $\boldsymbol{\hat{\sigma}}_\parallel$ and $\boldsymbol{\hat{v}}_\parallel$, respectively, and are solutions of the Stokes equations. The lubrication approximation is used here. The solution reads:
\begin{equation}
\label{eq:model-solution_lateral}
\hat{p}_\parallel(\boldsymbol{r}) = \frac{6\eta \hat{V}_\parallel r\cos\theta}{5\hat{h}^2(\boldsymbol{r})}, \quad \quad \boldsymbol{\hat{v}}_\parallel(\boldsymbol{r},z) = \frac{\boldsymbol{\nabla}\hat{p}_\parallel(\boldsymbol{r})}{2\eta} z(z-\hat{h}(\boldsymbol{r})) + \boldsymbol{\hat{V}}_\parallel \,\frac{z}{\hat{h}(\boldsymbol{r})}, 
\end{equation}
as shown in the main text. The Lorentz reciprocal theorem leads to:
\begin{equation}
\label{eq:Lorentz_lateral}
\boldsymbol{\hat{V}}_\parallel \cdot \boldsymbol{F} = \hat{V}_\parallel F_x = \int_{\mathcal{S}}   \boldsymbol{n} \cdot \boldsymbol{\hat{\sigma}}_\parallel \cdot \boldsymbol{v} \, \mathrm{d}s.
\end{equation}
Using the lubrication expression of the stress tensor of the model problem, $\boldsymbol{\hat{\sigma}}_\parallel \simeq -\hat{p}_\parallel\mathbf{I} + \eta \boldsymbol{e}_z \partial_z \boldsymbol{\hat{v}}_\parallel$, we get an expression for the tangential force as:
\begin{equation}
\label{eq:reciprocal_lateral}
F_x    = \frac{1}{\hat{V}_\parallel} \int_{\mathbb{R}^2} \bigg[ -\eta \partial_z \boldsymbol{\hat{v}}_\parallel \cdot u(t) \boldsymbol{e}_x  -\hat{p}_\parallel(\partial_t - u(t)\partial_x)\delta  - \eta (\partial_z \boldsymbol{\hat{v}}_\parallel \cdot  \partial_z\boldsymbol{v}_0) \delta \bigg] \textrm{d}\boldsymbol{r}.
\end{equation}
For the same reason as the one invoked in the main text, the zeroth-order tangential drag force (\textit{i.e.} the integral of $-\eta \partial_z \boldsymbol{\hat{v}}_\parallel \cdot u(t) \boldsymbol{e}_x$ in Eq.~\eqref{eq:reciprocal_lateral}) cannot be computed here as the integral diverges within the lubrication approximation. In contrast, the first-order EHD force is well defined in the lubrication framework and can be computed in Fourier space using Parseval's theorem, leading to:
\begin{equation}
\label{eq:reciprocal_lateral_result}
\kappa F_{x,1} = - \frac{3177 \pi ^3}{6400 \sqrt{2}} \frac{\eta^2 u \dot{d}(\lambda+2\mu)}{\mu(\lambda+\mu)} \bigg( \frac{a}{d}\bigg)^{5/2} + \frac{9 \pi ^3}{200 \sqrt{2}} \frac{\eta^2 \dot{u}a(\lambda+2\mu)}{\mu(\lambda+\mu)}\bigg( \frac{a}{d}\bigg)^{3/2}.
\end{equation}

\section{Thin compressible substrate}
\label{app:winkler}

\begin{figure}
\centering
\includegraphics[width=0.6\columnwidth]{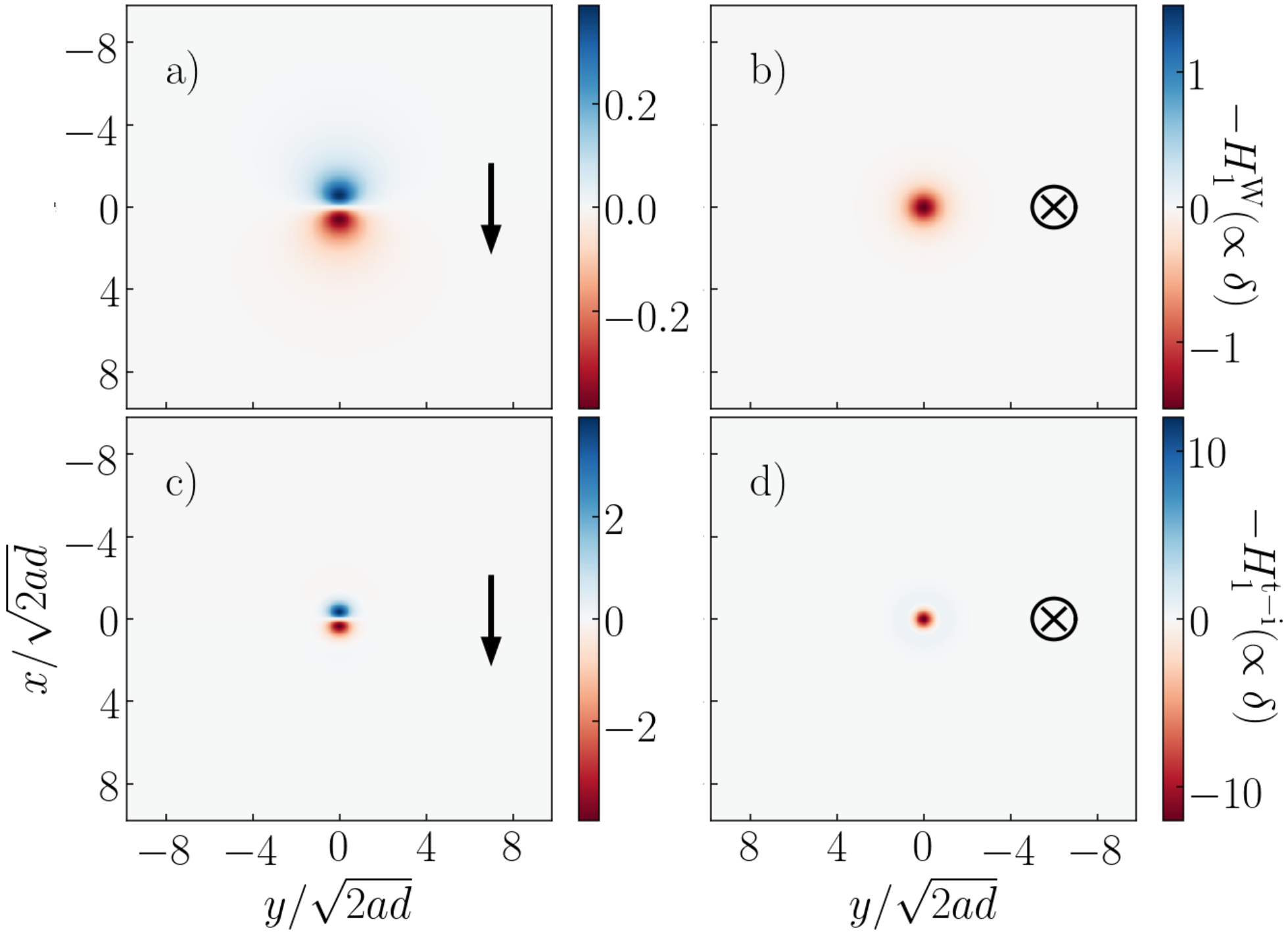}
\caption{Dimensionless deformation fields at the free surface of two soft substrates, for a sphere placed at a unit distance $D = 1$ and for two modes of motion. In a) and b), the substrate's mechanical response follows the Winkler foundation (see Eq.~\eqref{eq:Winkler}). In c) and d), the substrate's mechanical response is the one of a thin incompressible layer (see Eq.~\eqref{eq:thin-incomp}).}
\label{fig:H_1_thin}
\end{figure} 

In this appendix, we derive the EHD interactions exerted on a sphere immersed in a viscous fluid and near a thin compressible substrate of thickness $h_\textrm{sub}$. The deformation field follows the Winkler foundation:
\begin{equation}
\label{eq:Winkler}
\delta(\boldsymbol{r}, t) = -\frac{h_\textrm{sub}}{(2\mu + \lambda)} p(\boldsymbol{r},t),
\end{equation}
which is valid for substrates of thickness smaller than the typical extent of the pressure field, namely the hydrodynamic radius $\sqrt{2ad}$~\citep{leroy2011hydrodynamic,chandler2020validity,kargar2021lift}. We perform the same asymptotic expansion as the one in the main text, defining the Winkler dimensionless compliance as~\citep{salez2015elastohydrodynamics}: 
\begin{equation}
\kappa^\textrm{W} = \frac{\sqrt{2}h_\textrm{sub}\eta u^* a^{1/2}}{d^{*5/2} (2\mu + \lambda)}.
\end{equation}
The first-order substrate deformation, or equivalently here the zeroth-order pressure, reads:
\begin{equation}
H_1^\text{W}(\boldsymbol{R}, T) = P_0(\boldsymbol{R}, T) = \frac{3 \dot{D}}{2 (D+R^2)^2} + \frac{6 R U \cos\theta}{5(D+ R^2)^2}.
\end{equation}
The first-order deformation fields are plotted in Fig.~\ref{fig:H_1_thin}a) and b) for tangential and normal motions of the sphere, respectively. The deformation exhibits the same structure as the one in Fig.~\ref{fig:H_1} for semi-infinite substrates, but the lateral extent of the deformation is	narrower. This is expected as the deformation response induced by a given applied pressure is local for a thin compressible layer (see Eq.~\eqref{eq:Winkler}), while semi-infinite substrates display a non-local response due to the convolution of the pressure with their Green's function (see Eq.~\eqref{eq:elasticity_semiinf}). The first-order pressure correction follows the same type of equation as in the main text: 
\begin{equation}
\mathcal{L}.P_1^\textrm{W} = F_0^\textrm{W}(R,T) + F_1^\textrm{W}(R,T) \cos\theta + F_2^\textrm{W}(R,T) \cos2\theta,
\end{equation}
with:
\begin{equation}
F_0^\textrm{W}(R,T) = -\frac{144 R^2 U^2}{25 (D+R^2)^7}  \bigg[ D^2 - 6DR^2 +R^4   \bigg]  + \frac{18R^2 \dot{D}^2}{ (D+R^2)^7} \bigg[ 5D-4R^2 \bigg]  -\frac{18R^2 \ddot{D}}{ (D+R^2)^5},
\end{equation}
and:
\begin{equation}
F_1^\textrm{W}(R,T) = \frac{216 R^3 U\dot{D} }{5 (D+R^2)^7}  \bigg[ -5D + R^2   \bigg]  + \frac{72R \dot{U}}{5 (D+R^2)^5}.
\end{equation}
We note that $F_2^\textrm{W}$ does not contribute for the forces and torque. The isotropic component of the pressure can be found analytically, using polynomial fractions, as:
\begin{equation}
P_1^{\textrm{W},(0)}(R,T) =\frac{9}{125}\frac{7-5Y^2}{(1+Y^2)^5} \frac{U^2}{D^4} - \frac{3}{40}\frac{71+55Y^2+30Y^4}{(1+Y^2)^5} \frac{\dot{D}^2}{D^5} +\frac{3}{2}\frac{1}{(1+Y^2)^3} \frac{\ddot{D}}{D^4},
\end{equation}
where $Y = R/D^{1/2}$ is the self-similar variable. However, the first angular component of the pressure does not exhibit such an analytical solution, and is thus found by numerical integration of two scaling functions. Its general expression reads:
\begin{equation}
P_1^{\textrm{W},(1)}(R,T) =\frac{U\dot{D}}{D^{9/2}}\phi_{U\dot{D}}^\textrm{W}\bigg(\frac{R}{D^{1/2}}\bigg)+\frac{\dot{U}}{D^{7/2}}\phi_{\dot{U}}^\textrm{W}\bigg(\frac{R}{D^{1/2}}\bigg).
\end{equation}
Following the same calculation as in the main text, we find the normal force as:
\begin{equation}
\label{eq:EHD_normal_Winkler}
F_z^\textrm{W} = -\frac{6\pi \eta a^2 \dot{d}}{d} + \frac{48\pi}{125} \frac{\eta^2 u^2 h_\textrm{sub}}{a(2\mu +\lambda)} \bigg(\frac{a}{d}\bigg)^3 - \frac{72\pi}{5} \frac{\eta^2 \dot{d}^2 h_\textrm{sub}}{a(2\mu +\lambda)} \bigg(\frac{a}{d}\bigg)^4 + \frac{6 \pi \eta^2 \ddot{d} h_\textrm{sub}}{(2\mu +\lambda)} \bigg(\frac{a}{d}\bigg)^3.
\end{equation}
We stress that the prefactors $48\pi/125$ and $6\pi$ are in agreement with the results in~\cite{urzay2007elastohydrodynamic} and~\cite{leroy2011hydrodynamic}, respectively. Similarly, the force along $x$ reads:
\begin{equation}
\label{eq:EHD_lateral_Winkler}
F_{x}^\textrm{W} = 6 \pi \eta a u \bigg(\frac{8}{15}\log\left(\frac{d}{a}\right) - 0.95429 \bigg) - 23.9\frac{\eta^2u\dot{d}h_\textrm{sub}}{a(2\mu+\lambda)} \bigg(\frac{a}{d}\bigg)^3 + 4.520\frac{\eta^2\dot{u}h_\textrm{sub}}{(2\mu+\lambda)} \bigg(\frac{a}{d}\bigg)^2.
\end{equation}
The torque can be evaluated as well, and reads:
\begin{equation}
\label{eq:EHD_torque_Winkler}
T_{y}^\textrm{W} = 8\pi \eta u a^2 \bigg(-\frac{1}{10}\log\left(\frac{d}{a}\right) - 0.19296 \bigg) + 12.2 \frac{\eta^2u\dot{d}h_\textrm{sub}}{(2\mu+\lambda)} \bigg(\frac{a}{d}\bigg)^3 - 1.51\frac{\eta^2\dot{u}ah_\textrm{sub}}{(2\mu+\lambda)} \bigg(\frac{a}{d}\bigg)^2.
\end{equation}
All the prefactors for the EHD corrections of the tangential force and torque have been found numerically. Finally, following the approach in the main text, it is straightforward to generalize Eqs.~\eqref{eq:EHD_normal_Winkler},~\eqref{eq:EHD_lateral_Winkler}~and~\eqref{eq:EHD_torque_Winkler} in order to incorporate rotation.

\section{Thin incompressible substrate}
\label{app:thin-incompressible}
In this appendix, we suppose that the substrate of thickness $h_\textrm{sub}$ is incompressible, \textit{i.e.} of Poisson ratio $\nu = 1/2$, which means that the first Lam\'{e} coefficient $\lambda$ is infinite. In this situation, the Winkler foundation is not valid. Instead, the mechanical response of a thin substrate follows the relation (\cite{leroy2011hydrodynamic,chandler2020validity}):
\begin{equation}
\label{eq:thin-incomp}
\delta(\boldsymbol{r}, t) = \frac{h^3_\textrm{sub}}{3\mu} \boldsymbol{\nabla}^2 p(\boldsymbol{r},t), 
\end{equation}
where $\boldsymbol{\nabla}^2$ denotes the 2D Laplacian operator in the $(x,y)$ plan. We perform the same asymptotic expansion as in the main text, defining the thin-incompressible dimensionless compliance as: 
\begin{equation}
\kappa^\textrm{t-i} = \frac{\eta u^* h_\textrm{sub}^3 }{ 3\sqrt{2}\mu d^{*7/2} a^{1/2} }.
\end{equation}
The first-order substrate deformation reads:
\begin{equation}
H_1^\text{t-i}(\boldsymbol{R}, T) = -\boldsymbol{\nabla}^2 P_0(\boldsymbol{R}, T) = -\frac{12 \dot{D}(D-2R^2)}{ (D+R^2)^4} + \frac{48 R U (2D-R^2)\cos\theta}{5(D+ R^2)^4}.
\end{equation}
The deformation fields are plotted in Figs.~\ref{fig:H_1_thin} c) and d), for tangential and normal motions of the sphere, respectively. The first-order pressure correction follows the same type of equation as in the main text: 
\begin{equation}
\mathcal{L}.P_1^\textrm{t-i} = F_0^\textrm{t-i}(R,T) + F_1^\textrm{t-i}(R,T) \cos\theta + F_2^\textrm{t-i}(R,T) \cos2\theta,
\end{equation}
with:
\begin{equation}
\begin{split}
F_0^\textrm{t-i}(R,T)  &=\frac{1152 R^2 U^2 \left(R^2-2 D\right) \left(+2 D^2-11 R^2 D+2 R^4\right)}{25 \left(D+R^2\right)^9} \\
&+\frac{432 R^2 \left(2 D \left(D-5 R^2\right)+3 R^4\right) \dot{D}^2}{\left(D+R^2\right)^9}  +\frac{144 R^2 \left(2 R^2-D\right) \ddot{D}}{\left(D+R^2\right)^7},
\end{split}
\end{equation}
and:
\begin{equation}
F_1^\textrm{t-i}(R,T) =-\frac{2592 R^3 \dot{D}U \left(7 D^2-12 R^2 D+R^4\right) }{5 \left(D+R^2\right)^9}  -\frac{576 R^3\dot{U} \left(-2 D+R^2\right) }{5 \left(D+R^2\right)^7}.
\end{equation}
We note that $F_2^\textrm{t-i}$ does not contribute for the forces and torque. The isotropic component of the pressure can be found analytically, using polynomial fractions, as:
\begin{equation}
P_1^{\textrm{t-i},(0)}(R,T) =\frac{288 \left(7 Y^4-21 Y^2+17\right)}{875 \left(1+Y^2\right)^7}\frac{U^2}{D^5}  + \frac{126 Y^2-198}{7 \left(1+Y^2\right)^7} \frac{\dot{D}^2}{D^6} +\frac{36}{5 \left(1+Y^2\right)^5} \frac{\ddot{D}}{D^5},
\end{equation}
where $Y = R/D^{1/2}$ is the self-similar variable. However, the first angular component of the pressure does not exhibit such an analytical solution, and is thus found by numerical integration of two scaling functions. Its general expression reads:
\begin{equation}
P_1^{\textrm{t-i},(1)}(R,T) =\frac{U\dot{D}}{D^{11/2}}\phi_{U\dot{D}}^\textrm{t-i}\bigg(\frac{R}{D^{1/2}}\bigg)+\frac{\dot{U}}{D^{9/2}}\phi_{\dot{U}}^\textrm{t-i}\bigg(\frac{R}{D^{1/2}}\bigg).
\end{equation}
Following the same calculation as in the main text, we find the normal force as:
\begin{equation}
F_z^\textrm{t-i} = -\frac{6\pi \eta a^2 \dot{d}}{d} + \frac{432 \pi }{875} \frac{\eta^2 u^2 h_\textrm{sub}^3}{a^3\mu} \bigg(\frac{a}{d}\bigg)^4 - \frac{192 \pi}{35} \frac{\eta^2 \dot{d}^2 h_\textrm{sub}^3}{a^3\mu} \bigg(\frac{a}{d}\bigg)^5 + \frac{12 \pi }{5}\frac{ \eta^2 \ddot{d} h_\textrm{sub}^3}{a^2\mu} \bigg(\frac{a}{d}\bigg)^4.
\end{equation}
We stress that the prefactor $12\pi/5$ is consistent with the linear-response theory in~\cite{leroy2011hydrodynamic}. Similarly, the force along $x$ reads:
\begin{equation}
F_{x}^\textrm{t-i} = 6 \pi \eta a u \bigg(\frac{8}{15}\log\left(\frac{d}{a}\right) - 0.95429 \bigg) - 12.2\frac{\eta^2u\dot{d}h_\textrm{sub}^3}{a^3\mu} \bigg(\frac{a}{d}\bigg)^4 + 2.41\frac{\eta^2\dot{u}h_\textrm{sub}^3}{a^2\mu} \bigg(\frac{a}{d}\bigg)^3.
\end{equation}
The torque can be evaluated as well, and reads:
\begin{equation}
T_{y}^\textrm{t-i} = 8\pi \eta u a^2 \bigg(-\frac{1}{10}\log\left(\frac{d}{a}\right) - 0.19296 \bigg) + 7.75 \frac{\eta^2u\dot{d}h_\textrm{sub}^3}{a^2\mu} \bigg(\frac{a}{d}\bigg)^4 - 0.804\frac{\eta^2\dot{u}h_\textrm{sub}^3}{a \mu} \bigg(\frac{a}{d}\bigg)^3.
\end{equation}
Here again, the prefactors of the transverse force and torque are found using the Lorentz reciprocal theorem, as discussed above in the appendix. We stress that the thin-incompressible limit is mathematically valid for strictly incompressible substrates, but that its range of application is limited in practice. Indeed, usual elastomers and gels, that are considered as almost incompressible, have a Poisson ratio close to $\nu \simeq 0.49$, and thus a tiny but finite compressibility. In recent studies~\cite{Saintyves2016,rallabandi2017rotation,Saintyves2020}, it has been observed that the mechanical response of very thin incompressible elastic substrates is better described by the Winkler foundation (\textit{i.e.} thin and compressible) than the thin-incompressible limit discussed here. This observation has then been established on solid theoretical grounds for the EHD lift~\cite{chandler2020validity}, and is intimately rooted in the structure of the elastic Green's function. An empirical scaling, based on the numerical calculation of the EHD lift coefficient, has been derived subsequently~\cite{kargar2021lift} and suggests that the thin-incompressible model is valid for thicknesses comprised in the range $\frac{\sqrt{7}}{3}(1/2-\nu)^{1/2} \ll h_\text{sub}/\sqrt{2ad} \leq 0.12$. For $\nu = 0.49$, the lower bound of the latter range is $0.088$, which confirms that the validity window of the thin-incompressible model is limited. 

\bibliography{biblio}
\end{document}